\documentclass[]{interact}

\usepackage{epstopdf}
\usepackage[caption=false]{subfig}

\usepackage{graphicx}
\usepackage{dcolumn}
\usepackage{bm}

\usepackage[english]{babel}
\usepackage{float}
\usepackage{graphicx} 
\usepackage[export]{adjustbox}

\usepackage[numbers,sort&compress,merge]{natbib}
\bibpunct[, ]{[}{]}{,}{n}{,}{,}

\theoremstyle{plain}

\theoremstyle{definition}

\theoremstyle{remark}

\begin{document}


\title{Control of molecular ultracold plasma relaxation dynamics by mm-wave Rydberg-Rydberg transitions}

\author{
\name{Fernanda Banic Viana Martins\textsuperscript{a,b}, James S.~Keller\textsuperscript{a,b} and Edward R.~Grant\textsuperscript{b}\thanks{CONTACT E.~R. Grant. Email: edgrant@chem.ubc.ca}}
\affil{\textsuperscript{a} Department of Chemistry, Kenyon College, 
 Gambier, OH 43022; \\
 \textsuperscript{b} Department of Chemistry, and Department of Physics \& Astronomy,
 University of British Columbia, Vancouver, BC}
}

\maketitle

\begin{abstract}
Resonant mm-wave fields drive $n_0f(2) \rightarrow (n_0 \pm 1)g(2)$ transitions in a state-selected $n_0f(2)$ Rydberg gas of NO.  This transformation produces a clear signature in the selected field ionization spectrum and dramatically increases the intensity of corresponding features in the spectrum of the $n_0f(2)$ Rydberg series observed in UV-UV double resonant transitions via the $A ~^2\Sigma^+, ~N'=0$ state.  Here, $n_0$ refers to the principal quantum number of an $f$ Rydberg state converging to the $N^+=2$ rotational state of NO$^+ X~ ^1\Sigma^+$.  Enhancement owing to transitions from $\ell=3$ ($f$) to $\ell=4$ ($g$) appears both in the electron signal detected at early time by the field ionization of Rydberg molecules and, 40 $\mu$s later, as the late-peak signal of plasma in a state of arrested relaxation.  Similar stabilization and enhanced intensity also occurs for shorter-lived interloping complex resonances, $44p(0) - 43d(1)$  and  $43p(0) - 42d(1)$.  We conclude from these observations that avalanche alone does not guarantee a plasma state of arrested relaxation.  But rather, the formation of an arrested phase requires both avalanche-produced NO$^+$ ions and a persistent population of long-lived Rydberg molecules.  
\end{abstract}


\section{Introduction}

 With sufficient initial density, $\rho_0$, a high principal-quantum-number Rydberg gas of nitric oxide avalanches to form an ultracold plasma.  This plasma evolves in a process that quenches the electron temperature and divides population between a long-lived distribution of very weakly bound electrons and a decaying population of long-lived residual Rydberg molecules that retain the initial principal quantum number, $n_0$, with randomized angular momentum, $\ell \leq n_0$ \cite{Haenel2017}.  The resulting distribution of strongly coupled ions, electrons and Rydberg molecules expands and self-assembles to form a canonical state of temperature and density, far from conventional thermal equilibrium.  This quenched molecular ultracold plasma presents a unique laboratory in which to study the consequences of disorder in interacting quantum many-body systems \cite{Sous2018,SousNJP}. 

The properties of the arrested phase depend very little on the initial conditions of the Rydberg gas.  However, the {\it quantity} of a Rydberg gas that evolves to a long-lived plasma depends sensitively on the rate of neutral predissociation, which varies strongly with the penetration of the Rydberg orbital, as determined by its orbital angular momentum, $\ell$.   In a photoselection process confined to a final total angular momentum neglecting spin, $N=1$, only $\ell=3$ states converging to $N^+=2$ ($n_0f(2)$) have sufficient initial lifetime to undergo an avalanche and quench that support arrested relaxation.  

Rydberg-Rydberg transitions in the mm-wave region of the electromagnetic spectrum can populate states of higher orbital angular momentum.  By opening the door to initial states of $\ell > 3$, mm-wave excitation affords an avenue for the quantum-state control of molecular Rydberg gas relaxation dynamics.   

Rydberg states exhibit large cross sections for the absorption of mm-wave radiation owing to enormous dipole matrix elements for transitions between adjacent levels of high principal quantum number \cite{Gallagher2005,Tate2018}.  Approaches relying on mm-wave spectroscopy have played a crucial role in the precise measurement of Rydberg quantum defects \cite{Li2003,Schafer_Kr,Schafer_Xe}, the local determination of electrostatic fields \cite{Osterwalder1999}, the control of coherent population transfer \cite{Grimes2017} and the creation of dark state polaritons \cite{Tanasittikosol2011}.  Tuned mm-wave fields have been demonstrated to suppress predissociation, a nonradiative decay mechanism that has challenged the spectroscopic study of Rydberg molecules \cite{Murgu2001}.  Resonant frequencies can drive transitions to longer-lived states of higher $\ell$, stabilizing a molecular Rydberg gas with respect to neutral predissociation.  

Here, we show that mm-wave radiation, tuned to drive $n_0f(2) \rightarrow (n_0 \pm 1)g(2)$ transitions, act to alter the short-time dynamics by which NO Rydberg molecules avalanche to plasma, with consequences for the extent of arrested relaxation.  This early transformation to a Rydberg gas of higher $\ell$ dramatically increases the fraction of molecules in the plasma that form a persistent $n_0$ Rydberg signal in selective field ionization (SFI) measurements hundreds of nanoseconds later, and doubles the yield of plasma that survives as an arrested phase for 40 $\mu$s.

\section{Experimental}

\subsection{Production of a cold Rydberg gas of nitric oxide in a seeded supersonic molecular beam}

A molecular beam of nitric oxide  enters a vacuum chamber in a 1:10 supersonic beam of argon from a stagnation pressure of 5 bar. This jet travels 35 mm to transit a 1 mm diameter skimmer separating the source and experimental chambers of a differentially pumped vacuum system. 
\begin{figure}[h!]
\centering
\includegraphics[scale=0.6]{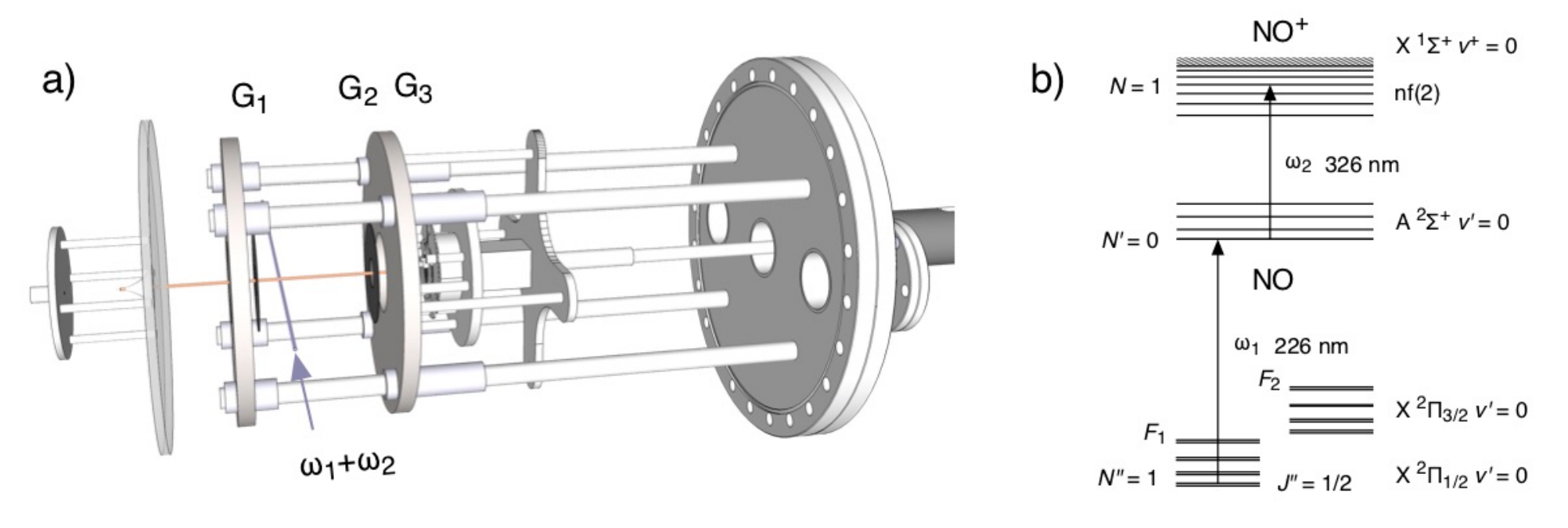}
\caption{(a) Co-propagating laser beams, $\omega_1$ and $\omega_2$, cross a molecular beam of nitric oxide between entrance aperture G$_1$ and grid G$_2$ of a differentially-pumped vacuum chamber. (b)  Nd:YAG-pumped dye lasers ($\omega_1$ and $\omega_2$)  pump ground state NO first to the excited $A\ ^2 \Sigma^+ \ N^\prime=0$ state and then to a Rydberg level with $N=1$. }
\label{fig:diagram}
\end{figure}

As shown in Fig.~\ref{fig:diagram}a, two co-propagating laser pulses that originate from a pair of frequency-doubled, Nd:YAG-pumped dye lasers intercept the NO beam between grids G$_1$ and G$_2$, converting it to a dense Rydberg gas. The first laser, $\omega_1$, excites  molecules from the ground state, $X \ ^2\Pi_{1/2} \ N ^{\prime \prime}=1$, to an intermediate state, $A\ ^2 \Sigma^+ \ N^\prime=0$. This assures that the absorption of the second laser, $\omega_2$, promotes excited molecules to selected Rydberg states of initial principal quantum number $n_0$, with total angular momentum neglecting spin constrained to $N=1$ (Fig. \ref{fig:diagram}b).  

\subsection{Evolution from Rydberg gas to ultracold plasma}

In the limit of saturated $\omega_1$ excitation, laser-crossed molecular beam illumination creates an ellipsoidal volume of Rydberg gas with a peak density in excess of $1\times 10^{12} \ \textrm{cm}^{-3}$ ($1~\mu{\rm m}^{-3}$) \cite{MSW_tutorial}.  At this density, and even many orders of magnitude lower, Rydberg-Rydberg interactions inevitably drive a Rydberg gas ensemble formed in a beam, or even in a magneto-optical trap (MOT), to plasma \cite{Killian_Phys_rept}.  It is just a matter of time.  For example, we have shown that a $5p^5 ~[2P ] ~6p ~^2[5/2]_2$ Rydberg gas of Xe with a density as low as $10^8~{\rm cm}^{-3}$ evolves to plasma on the timescale of 40 $\mu$s \cite{Hung2014}.  Coupled-rate simulations predict that a nitric oxide Rydberg gas with a density of $10^{12}~{\rm cm}^{-3}$ avalanches in less than 100 ns \cite{Saquet2012}.  Measurements in our laboratory have confirmed this timescale \cite{Haenel2017}.  For the purposes of the present experiment, we adjust the power of $\omega_2$ to produce an initial Rydberg gas with a density around $10^{11}~{\rm cm}^{-3}$, creating a system that evolves in 500 ns to form a mixture of ultracold plasma consisting of electrons bound by the space charge or single NO$^+$ ions in very high-$n$ Rydberg orbitals.  

With few exceptions among all the $N=1$ Rydberg states accessible in transitions from $A\ ^2 \Sigma^+ \ N^\prime=0$, only those in the $n_0f$ series converging to NO$^+$, $X\ ^1\Sigma^+$, $N^+ = 2$ limit have sufficient lifetime to evolve to plasma.  Tuning the frequency of $\omega_2$ determines which high-Rydberg state in this series it populates.  

\subsubsection{Field ionization as a gauge of short-time dynamics}

We gauge the survival of Rydberg molecules in the first few hundred nanoseconds by applying a forward-biased (negative) voltage ramp to G$_1$ that begins 300 ns after $\omega_2$ and rises at a rate of 0.8 V ns$^{-1}$.  This rising field separates electrons from the NO$^+$ space charge and ionizes very high-$n$ Rydberg molecules.  These electrons appear as a prompt peak at low field.  When the ramp reaches 100 V cm$^{-1}$ or so (the precise value depends upon $n_0$, the initial principal quantum number), a rising electron signal marks the selective field ionization (SFI) of an enduring population of $n_0f(2)$ Rydberg molecules.  

The full SFI ramp collects all of the extravalent electrons in a given plasma volume \cite{Haenel2017}.  We calibrate this signal, determined for a given ramp delay, as a relative gauge of $\rho_0$, the initial density of the Rydberg gas, by making measurements in which we precisely delay the second laser pulse, $\omega_2$, in our $\omega_1+\omega_2$ sequence of double-resonant excitation.  Here, the decay of the $A ~^2\Sigma^+$ state of NO with a well-known lifetime \cite{Astate} enables us to diminish $\rho_0$ to a precisely controlled degree and calibrate the total field ionization signal for initial density.  

For present purposes, we start the SFI ramp 300 ns after pulsing $\omega_2$, and collect the electron signal produced approximately 200 ns later when the field rises to ionize the $n_0$ Rydberg molecules that remain for a fixed range of $\rho_0$.  We monitor the change in amplitude and appearance potential of this signal in the presence of resonant mm-wave radiation to determine the effect of early time Rydberg-Rydberg transitions on the initial progress of an ultracold plasma to a long-lived state.  

\subsubsection{Field-free propagation and long-time durability}

In the absence of a field ramp, the volume of photoexcited gas travels with the laboratory velocity of the molecular beam for an elapsed time of 40 $\mu$s to reach a grid, G$_2$, that caps the field-free region.  There, it encounters a static field applied between grids G$_2$ and G$_3$, set typically to 200~V/cm. The electron signal produced as the excited volume transits G$_2$ forms a late peak that traces the width of the evolving plasma in the direction of propagation.  As in the SFI experiment, integrating this electron signal waveform measures all of the extravalent electrons that are present in the illuminated volume.  

We vary the  $\omega_1- \omega_2$ delay to regulate the average initial density of the Rydberg gas.  For a controlled density, we measure the change in the integrated late-peak signal in the presence of resonant mm-wave radiation to gauge the effect of early time Rydberg-Rydberg transitions on the yield of plasma that evolves to a long-time state of arrested relaxation.  

\subsection{UV-UV double-resonance and UV-UV-mm-wave triple-resonance spectroscopy}

Detecting either the $n_0f(2)$ Rydberg molecules resolved by selective field ionization, or the late-peak ultracold plasma signal, we scan $\omega_2$ to obtain an action spectrum of Rydberg resonances descending to 100 cm$^{-1}$ below the lowest ionization threshold of NO.  With SFI detection, this spectrum marks Rydberg molecules that survive the 300 ns ramp-field delay.  A late-peak $\omega_2$ scan maps the spectrum of initial states of Rydberg gas that form a long-lived ultracold plasma.  

We perform the same two types of scans in the presence of CW, 100 $\mu$W cm$^{-2}$ mm-wave fields of selected frequency.  This tests whether a Rydberg-Rydberg transition driven at that frequency affects the fraction of the Rydberg gas that survives to yield an electron signal, after either a short delay of 300 ns or a long flight time of 40 $\mu$s.

Fixing $\omega_1$ on the $A~^2\Sigma^+ ~v'=0~N'=0$ state, and $\omega_2$ on a selected resonance in the high-Rydberg spectrum, we scan the mm-wave frequency as described below.  The variation in electron signal, detected either by selective field ionization or by the electrons collected in the late peak, traces the lineshapes of high principal quantum number Rydberg-Rydberg transitions, and describes the effect of these transitions on avalanche dynamics - written in their effect on short- and long-term Rydberg/plasma lifetime.  

Even small electrostatic fields formed by surface charge in an apparatus and unshielded ions in the sample can act to perturb the energy and other properties of high Rydberg states.  For that reason, we have made a determined effort to minimize stray fields to the range of tens of millivolts.  We have tested the effects of applied DC fields from $\pm 100$ mV to $\pm 2$ V, finding that the presence of a DC field significantly affects the intensity of resonances, but perturbs linewidths only slightly \cite{Wang_mm}.  We have used the forward/reverse-bias symmetry of these effects to minimize stray fields between G$_1$ and G$_2$.

\subsection{Production and control of mm-wave radiation}

We operate a Virginia Diodes 10 - 400 GHz solid-state multi-band transmitter (VDI-Tx-S129) to produce narrow-bandwidth mm-wave radiation of selected frequency in a range from 60 to 120 GHz.  This system operates with three primary low-frequency sources:  (1) An on-board, computer-controlled Micro Lambda MLSE-1006 yttrium iron garnet (YIG) oscillator-based frequency synthesizer provides precise fixed frequencies from 10 to 20 GHz;   (2) A programmable DC power supply (Tektronix PWS4721) drives a Virginia Diodes voltage-controlled oscillator (VCO) system (VDI-Tx-S130) to generate time-varying primary fields of frequency from 8 to 20 GHz;  (3) The programmed output of a Tektronix AWG7102 10 GS/s arbitrary waveform generator creates pulsed carrier waveforms with mm-wave frequencies from 1.9 to 7.7 GHz for conversion to output from 12.5 to 18.3 GHz in a VDI-MixAMC-S116 Sideband Generator, which then serves as a primary input.  

The multi-band transmitter, operating in band 5, frequency doubles and triples the low-frequency input, and applies a gain of 5.  It broadcasts the output to free space via a VDI-WR-10 $35.5 \times 16.3$ mm conical horn.  A 50 mm diameter teflon lens with a nominal focal length of 300 mm collects this mm-wave radiation and focuses it through a silica window to cross the molecular beam, counter-propagating with the combined unfocused $\omega_1+\omega_2$ laser beam.  In band 5, which extends from 70 to 110 GHz, the multi-band transmitter outputs a power of 14 dBm (25 mW), which yields a power density at the molecular beam of about 100 $\mu$W cm$^{-2}$.

\section{Results}

\subsection{Double-resonant excitation spectra of high-Rydberg states}

A doubled, Nd:YAG-pumped dye laser, tuned to 226 nm with a collimated FWHM of 1 mm and a pulse energy of $\sim 3 ~\mu$J ($\omega_1$) intersects the seeded supersonic molecular beam to define a 2 mm$^3$ ellipsoidal volume element.  Here, a rotationally resolved $^PQ_{1}(0.5)$ $N''=1$ $\gamma$-band transition excites  NO molecules to the $A ~^2 \Sigma^+ ~ N'=0$ state.  These angular-momentum-selected $A$-state molecules absorb light from a second Nd:YAG-pumped dye laser pulse ($\omega_2$), tuned from 330 to 327 nm with a pulse energy of 7 mJ, to form an excitation spectrum of transitions to high Rydberg states.  Varying the $\omega_1 -\omega_2$ delay controls the initial Rydberg gas density, which for results presented here we estimate to be $\sim 10^{11}$ cm$^{-3}$  ($ 0.1~{\mu \rm m}^{-3}$).  

\begin{figure}[h!]
\centering
\includegraphics[scale = 0.75]{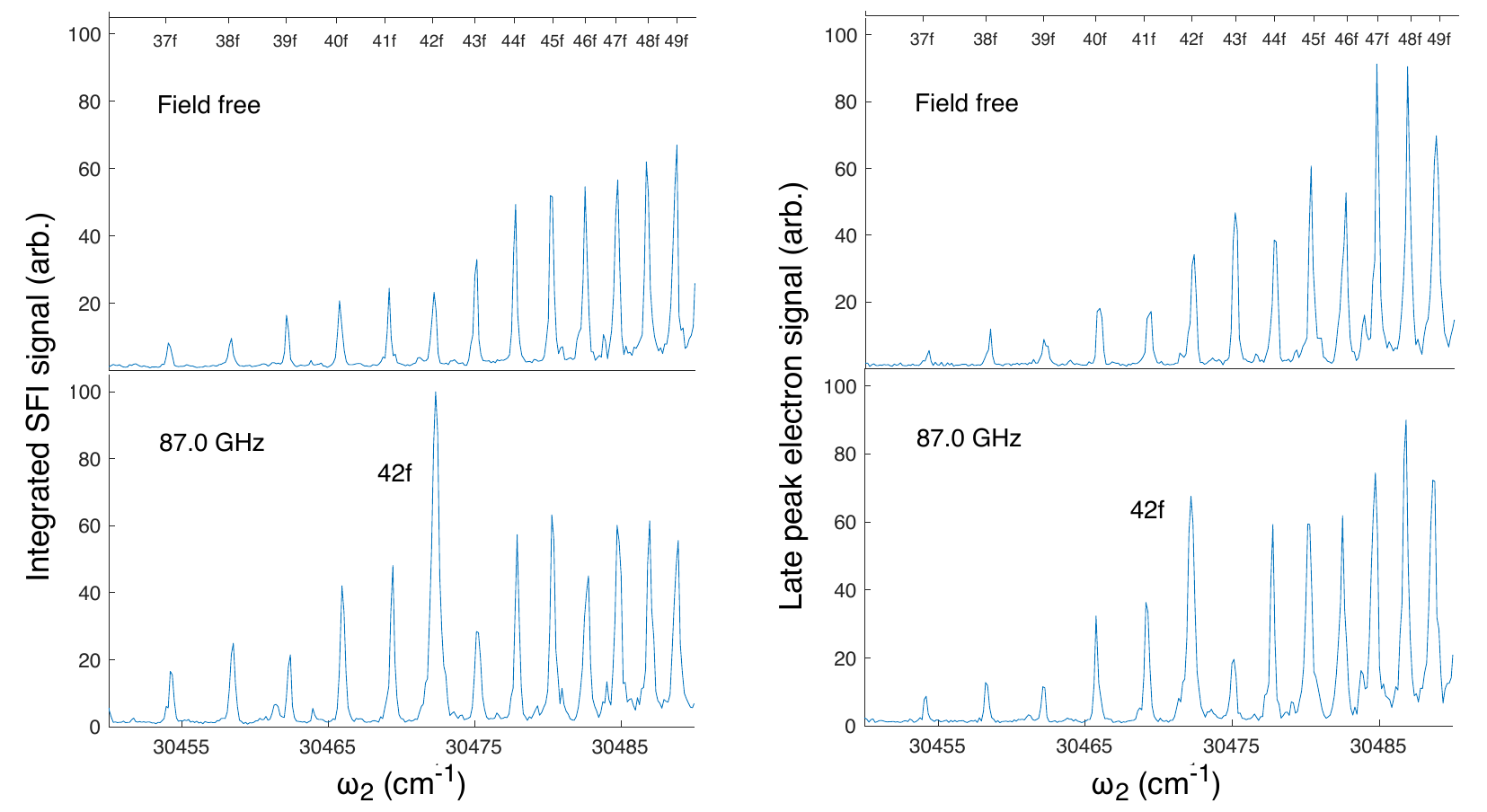}
\caption{Double-resonant $\omega_2$ spectra of nitric oxide showing resonances in the $n_0f(2)$ series for $n_0$ from 37 to 49.  (left) Electron signal integrated over Rydberg resonances in the SFI spectra. (right) Electron signal integrated over the late peak after a flight time of 40 $\mu$s.  (top frames) Field-free conditions.  (bottom frames) In the presence of a 100 $\mu$W cm$^{-2}$ CW mm-wave field with a frequency of 87.0 GHz. }
\label{fig:n42}
\end{figure}

Figure \ref{fig:n42} shows a pair of typical $\omega_2$ excitation spectra.  Here -- in the signal detected by SFI, with a nominal delay of 300 ns, or, after a flight time of 40 $\mu$s, in the electron signal extracted by an electrostatic field applied between G$_2$ and G$_3$ -- we see a spectrum of the $N=1$ Rydberg states in a single $\ell = 3$ series converging to the $N^+=2$ rotational level of the $X ~^1 \Sigma^+$ NO$^+$ ion.   This $n_0 f(2)$ series appears in isolation despite the existence of a great many, fully allowed transitions to numerous other series with total angular momentum neglecting spin, $N$, equal to 1.   

The application of a mm-wave field at any one of a number of particular frequencies in the range from 75 to 110 GHz affects this pattern of $n_0 f(2)$ intensities.  Changes occur to a similar degree in $\omega_2$ excitation spectra detected either by integrating the prompt SFI signal or waiting 40 $\mu$s and collecting the late peak of electron signal formed as the illuminated volume transits G$_2$.  

The lower frames of Figure \ref{fig:n42} show SFI and late-peak $\omega_2$ excitation spectra obtained in the presence of a 100 $\mu$W cm$^{-2}$ mm-wave field with a frequency of 87.0 GHz.  Note the marked increase in the relative intensity of the $42f(2)$ feature, which occurs in both the prompt SFI and delayed late-peak excitation spectra.  Note also that the presence of the 87.0 GHz mm-wave field suppresses the intensity of the $43f(2)$ resonance in both cases.    

\begin{figure}[b!]
\centering
\includegraphics[scale = 0.67]{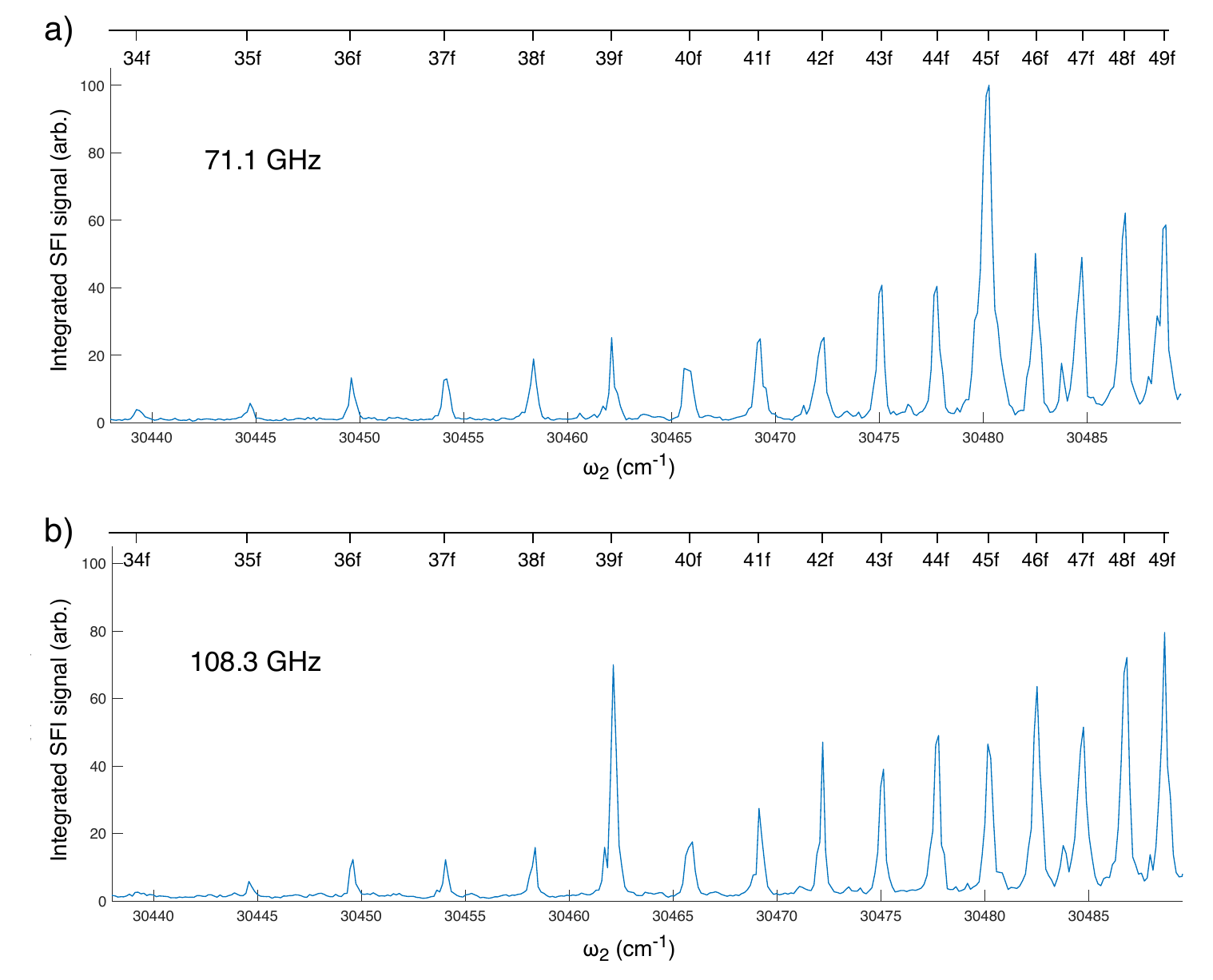}
\caption{Resonances in $\omega_2$ excitation spectra from $34f(2)$ to $49f(2)$, detected in the integrated SFI signal in the presence of a 100 $\mu$W cm$^{-2}$ mm-wave field tuned to  a) 71.1 GHz and b) 108.3 GHz.}
\label{fig:n39PFI}
\end{figure}

Other resonances in the $\omega_2$ spectrum from $n_0 = 34$ to 49 show similar patterns of enhancement and suppression caused by mm-wave fields of other specific frequencies.  For example, the top $\omega_2$ excitation spectrum in Figure \ref{fig:n39PFI} shows that a 100 $\mu$W cm$^{-2}$ mm-wave field tuned to 71.1 GHz increases the intensity of the $45f(2)$ resonance detected in the integrated SFI signal.  On the bottom, we see that a mm-wave field with a frequency of 108.3 GHz increases the intensity of the $39f(2)$ resonance. 

\begin{figure}[h!]
\centering
\includegraphics[scale = 0.8]{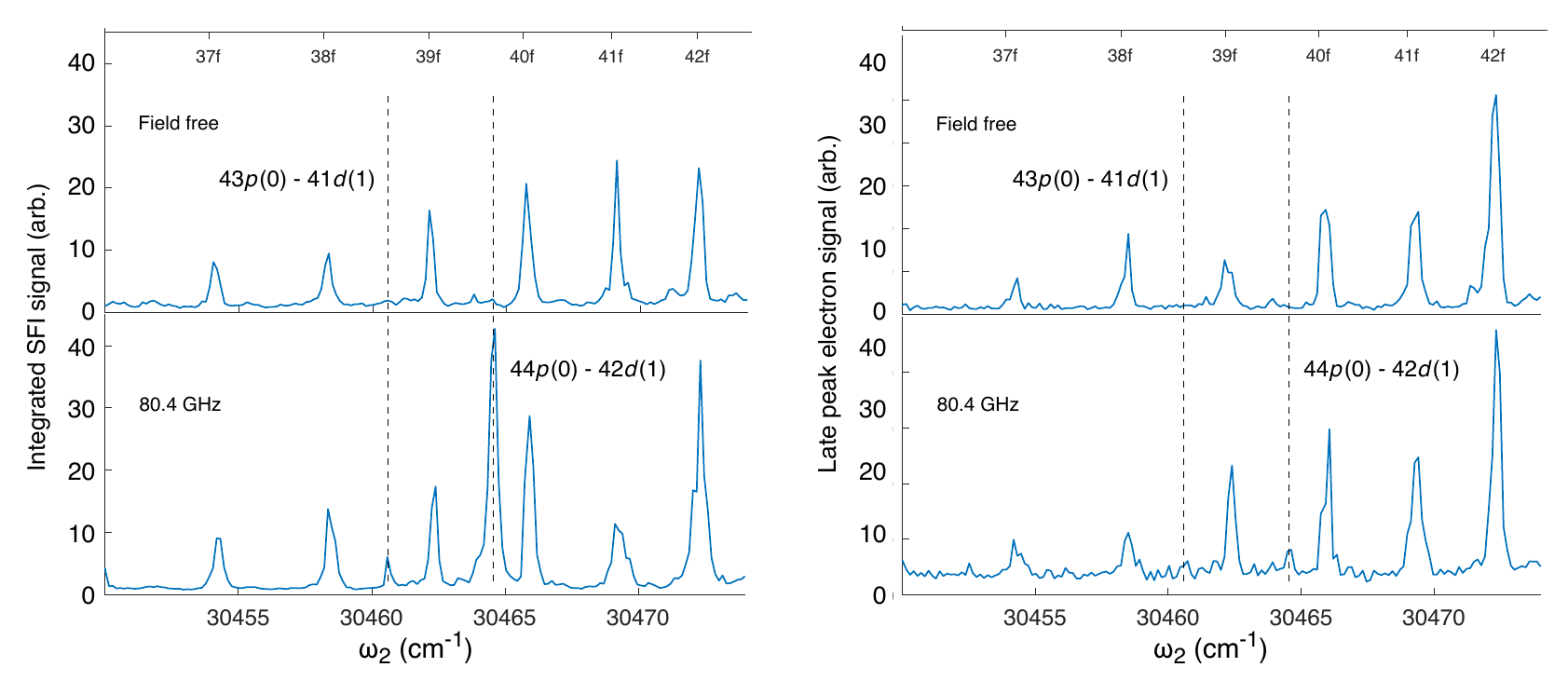}
\caption{Double-resonant $\omega_2$ spectra of nitric oxide showing resonances in the $n_0f(2)$ series for $n_0$ from 37 to 42. (left) Electron signal integrated over Rydberg resonances in the SFI spectra and (right) electron signal integrated over the late peak after a flight time of 40 $\mu$s. (top frames) Field-free conditions. (bottom frames) In the presence of a 100 $\mu$W cm$^{-2}$ CW mm-wave field oscillating at 80.4 GHz.  }
\label{fig:n44p}
\end{figure}

Figure \ref{fig:n44p} shows a special case of mm-wave enhancement in the $\omega_2$ spectrum.  Here, under conditions of low initial Rydberg gas density ($< 0.1~{\mu \rm m}^{-3}$), we find that a continuous mm-wave field with a frequency of 80.4 GHz depresses the $41f(2)$ resonance, while producing an entirely new feature in the spectrum between $38f(2)$ and $39f(2)$, and, with much greater intensity, between $39f(2)$ and $40f(2)$.  We detect these interloping resonances as prominent features at Rydberg appearance potentials in the SFI spectrum.  They also give rise to a long-lived plasma observed weakly in the $\omega_2$ excitation spectrum detected after 40 $\mu$s in the late-peak signal.

  \begin{figure}[h!]
\centering
\includegraphics[scale = 0.56]{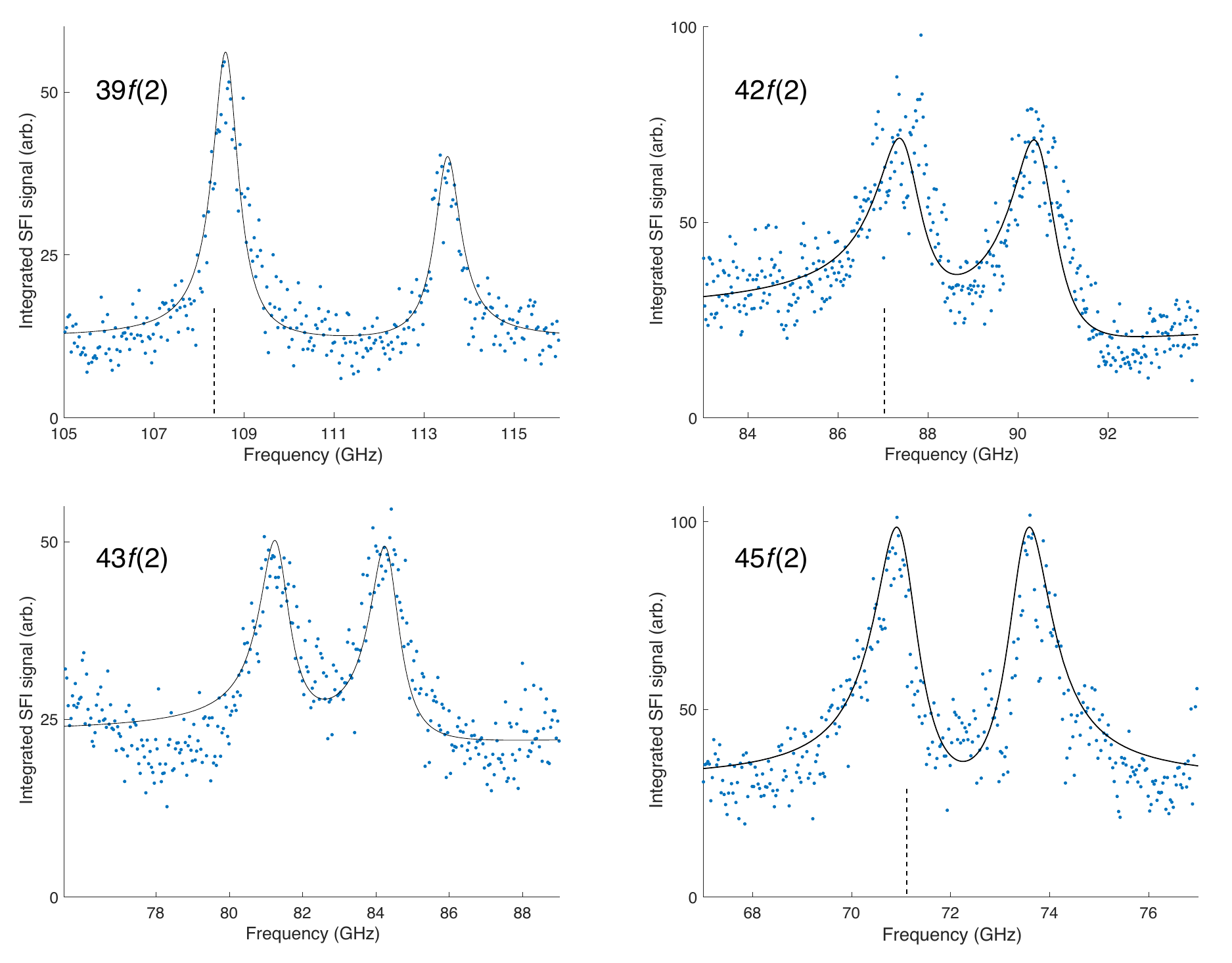}
\caption{Millimeter-wave excitation spectra detected in the signal of surviving Rydberg molecules field-ionized by an SFI ramp begun 300 ns after $\omega_2$ for initial $n_0f(2)$ Rydberg states, $n_0 = 39$, 42, 43 and 45.  Dashed lines mark the mm-wave frequencies chosen to enhance the intensity of selected resonances in Figures \ref{fig:n42} and \ref{fig:n39PFI}. }
\label{fig:n42mm}
\end{figure}

 \subsection{mm-wave excitation spectra of state-selected $nf(2)$ Rydberg gases}

Rydberg gases prepared in single initial quantum states exhibit distinctive mm-wave excitation spectra, detected in the integrated SFI spectrum sampled 300 ns after laser pulses, $\omega_1 + \omega_2$, and similarly in the late peak after a field-free flight time of 40 $\mu$s.   Figure \ref{fig:n42mm} displays spectra observed in the integrated SFI yield for Rydberg gases prepared with initial states, $39f(2)$, $42f(2)$ $43f(2)$ and $45f(2)$, respectively.  

 Here we see broad features signifying an enhanced SFI signal as we tune the mm-wave field through the resonant frequencies for transitions from $n_0f(2)$ to  $(n_0+1)g(2)$ and $ (n_0-1)g(2)$.  Measuring stepwise attenuation of the mm-wave field by means of neutral density filters, we establish that the transitions represented here occur in a linear regime of Rydberg-Rydberg absorption.  The intensities of these features affirm that for an initial density, $\rho_0=0.1~\mu$m$^{-3}$, the presence of a resonant mm-wave field increases the Rydberg signal detected in the SFI spectrum by about a factor of two.

 \subsection{Selective Field Ionization spectra in the presence of a mm-wave field}

\begin{figure}[h!]
\centering
\includegraphics[scale = 0.47]{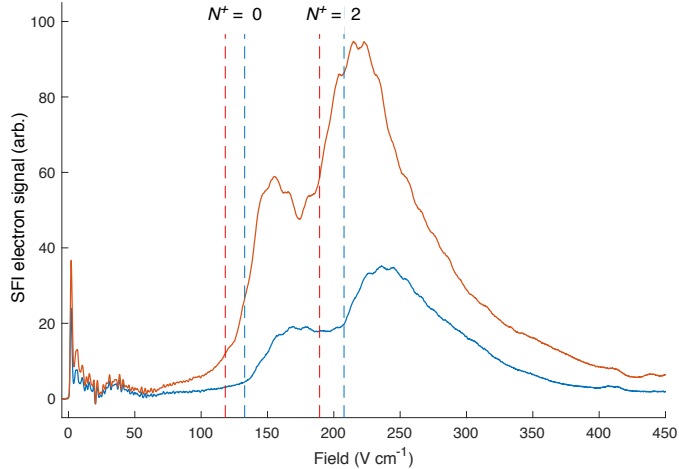}
\caption{(colour online) SFI spectrum of a $43f(2)$ Rydberg gas of NO obtained with a ramp delay of 300 ns.  (lower blue curve) Field ionization of $43f(2)$ to form NO$^+$ $N^+=0$ and 2.  Diabatic field ionization thresholds marked by blue dashed lines.  (upper red curve)  SFI spectrum of the same $43f(2)$ Rydberg gas obtained after applying a 100 ns wide pulsed 81.3 GHz mm-wave field resonant with the $43f(2) \rightarrow 44g(2)$ transition.  Here we see the field ionization of $44g(2)$ to form NO$^+$ $N^+=0$ and 2.  Diabatic field ionization thresholds marked by red dashed lines.  }
\label{fig:SFI}
\end{figure}

SFI spectra, such as the ones shown in Figure \ref{fig:SFI}, map the spectrum of electron binding energies in a Rydberg gas \cite{GallagherRA}.  These spectra show two clear signatures.  A prompt signal extracted at low voltage corresponds to electrons of low binding energy.  At higher field, each spectrum displays a pair of features with the same appearance potential difference.

This appearance potential difference corresponds in each case to the energy spacing of the NO$^+$ rotational states, $N^+=0$ and 2.  This characteristic property of the SFI spectrum of NO forms as a rising field acts on a single selected Rydberg state, driving it in a near-diabatic limit through the $n\ell (2)$ Stark manifold \cite{Fielding}.  As this wavepacket crosses the lowest field ionization threshold, the evolving population couples with the $N^+=0$ continuum, producing the electron signal observed as the lower-field feature in each case \cite{Haenel2017}.  

Here we also see that a pulsed 81.3 GHz mm-wave field with a duration of 100 ns applied with $\omega_2$ before the beginning of a delayed field ramp affects the early time quantum state distribution of a Rydberg gas prepared to occupy the initial state, $43f(2)$.  This resonance with the $43f(2) \rightarrow 44g(2)$ transition shifts the Rydberg signal to a lower appearance potential, signifying a decrease in binding energy.  Note again that a resonant mm-wave field with a peak power of 100 $\mu$W cm$^{-2}$ roughly doubles the intensity of the SFI signal.

 \section{Discussion}
 
\subsection{Background}

The molecular ultracold plasma of nitric oxide shows evident effects of mm-wave radiation in ($i$) $\omega_2$ spectra of relatively prompt and long-delayed electron signals, ($ii$) in the structure of the selective field ionization spectrum, and ($iii$) as SFI-detected resonances in the mm-wave excitation spectrum of selected high-Rydberg states.  The mm-wave transitions involved in each case cause factor-of-two changes in ultracold plasma yields.

\subsubsection{Definition of plasma yield}

This work uses two relative measures of plasma quantity.  The rising electrostatic field in a selective field ionization (SFI) measurement collects all of the extravalent electrons in a given plasma volume.  As discussed above, we have  calibrated this signal, in the absence of a mm-wave field, as a relative gauge of $\rho_0$, the initial density of the Rydberg gas. 

On a longer timescale, we allow the illuminated volume, propagating with the molecular beam, to pass through a grid, G$_2$, that caps the DC field-free region.  There, the plasma encounters a high electrostatic field that similarly extracts all of its extravalent electrons.  We vary the  $\omega_1- \omega_2$ delay to determine a late-peak signal as it relates to $\rho_0$.  As described above, we know $\rho_0$ as an absolute quantity from the known density of NO as a function of propagation distance in our molecular beam, together with calibrated efficiencies of $\omega_1$ and $\omega_2$ photoexcitation.

\subsubsection{Electron binding in the molecular ultracold plasma of nitric oxide}

The SFI spectrum resolves NO$^+$ ions that bind electrons to a recognizable degree defined by a quantity, $R/n^{*2}$, where $n^{*} = n_0 - \delta$, refers to the effective principal quantum number of a particular Rydberg state with quantum defect, $\delta$.  It clearly distinguishes this $n_0$-Rydberg signal from that of NO$^+$ ions that individually or collectively bind ions by energies of a few hundred GHz.  As this ensemble approaches a state of arrested relaxation, both of these populations appear to become anomalously durable.  

We cannot describe this system simply as a stable, very-high principal-quantum-number Rydberg gas.  It exhibits too great a density over the first 10 $\mu$s to sustain individual Rydberg molecules, each in an isolated central-field limit.  Well-calibrated semi-classical models call instead for such a gas to promptly decay by Penning ionization \cite{Robicheaux05}, followed by rapid avalanche \cite{Kuzmin:2002,Robicheaux:2002,Scaling}.  

Nor, can we assign this low-field band in the SFI spectrum to a low-temperature gas of free electrons weakly bound by a space charge of NO$^+$ ions.  If such were the case, well-known cross sections for three-body recombination would open a channel of Rydberg relaxation that would rapidly increase the electron temperature \cite{PVS}.  

Conventional notions of lower-$n$ Rydberg molecules fail to account for the presence of long-lived signals at higher appearance potentials in the SFI spectrum.  Any dynamic state of a system of NO$^+$ ions and weakly bound electrons would evolve in an $\ell$-mixing collisional environment \cite{Raithel04}, in which even the least penetrating of lower-$n$ Rydberg molecules would dynamically sample orbital angular momentum configurations that rapidly predissociate (see below) \cite{Saquet2012}.  

Nevertheless, these questions associated with the description of the ultimate plasma state of arrested relaxation do not affect the interpretation of the mm-wave resonances observed here.  It is sufficient to say, as observed by Gallagher and coworkers \cite{Murgu2001}, that mm-wave transitions serve to shelve population in longer-lived less-penetrating neutral states, and that suppressed predissociation during the initial phase of plasma evolution somehow increases the magnitude of particular electron signal waveforms detected later.

\subsubsection{Predissociation in the Rydberg states of nitric oxide}

Owing to its molecular character, the NO$^+$ - e$^-$ system thus differs significantly from the atomic ultracold plasma formed for example in a magneto-optical trap (MOT) \cite{Killian_Phys_rept}.  The ionization threshold of NO exceeds the bond dissociation energy to form N($^4{\rm S}$) + O($^3{\rm P}$) by nearly a factor of two.  Rydberg predissociation, ${\rm NO^*} (n,\ell) \rightarrow {\rm N}(^4{\rm S}) + {\rm O}(^3{\rm P})$, figures significantly in plasma avalanche dynamics \cite{Sadeghi:2011} and the development of spatial correlation \cite{Sadeghi:2014}.  

The rate at which a given Rydberg state of nitric oxide predissociates depends sensitively on its principal and orbital angular momentum quantum numbers, $n$ and $\ell$ \cite{Bixon,Murgu2001} .  Generally speaking, for a given $\ell$, predissociation rates fall with increasing $n$ by $1/n^3$.  Penetrating states of low $\ell$ decay much faster than states of high orbital angular momentum.  

For example, under field-free conditions, Vrakking and Lee \cite{Vrakking1995} measured a time constant, $\tau$, of approximately 400 ps for the $50p(0)$ state of NO, consistent with a trend for the series, $\tau_p = 2 \pi n^3/1.9 \times 10^{15}$ s.  Similar measurements found lifetimes forty times longer for $nf(2)$ states, described for the series by $\tau_f = 2 \pi n^3/5 \times 10^{13} ~{\rm s}$.  We note that with few exceptions, the only features in the ultracold plasma $\omega_2$ excitation spectrum of transitions from intermediate $A ~^2 \Sigma^+ ~ N'=0$ consist of $N=1$ states in the long-lived $nf(2)$ series, whether detected by SFI or as an ultracold plasma late peak.    

Vrakking and Remacle \cite{Remacle1998} determined that electrostatic fields lengthen the predissociation lifetimes of $n,\ell$-selected Rydberg states of NO to a degree consistent with an expected mixing of the low-$\ell$ state in a widening Stark manifold of high-$\ell$ states.  During the course of avalanche to plasma in our experiment, the SFI spectrum shows evidence that electron collisions also act to $\ell$-mix the selected $n_0f(2)$ state of the Rydberg gas initially formed under DC field-free conditions \cite{Haenel2017}, and we believe that this contributes essential short-term stability to the evolving system.  

Gallagher and coworkers \cite{Murgu2001} have shown that mm-wave radiation tuned to resonance with selected $n_0f \rightarrow (n_0 + 1)g$ and  $n_0f \rightarrow (n_0 - 1)g$ transitions can also serve to stabilize $nf$ Rydberg states of NO.  For example, in the presence of mm-wave fields of 48.92 and 50.40 GHz, double-resonant excitation tuned to the $51f(2)$ state with a lifetime of 16 ns \cite{Vrakking1995}, produces a field ionization signal for a 630 V/cm pulse applied 750 ns after photoexcitation.  These frequencies and the extended lifetime are well-explained by mm-wave transitions from $51f(2)$ to long-lived $50g(2)$ and $52g(2)$, much as reported here for $n_0f(2)$ Rydberg gases evolving to ultracold plasmas.

\subsubsection{Selective field ionization as a probe of Rydberg relaxation dynamics}

In the SFI measurement, a ramped electrostatic field couples crossing Stark levels, and drives Rydberg electron wavepackets into accessible continua.  In a molecule, the Stark mixing of rotational and electron orbital angular momentum complicates this picture.  The point in a rising field at which electrons first appear in the continuum of a given ion rotational state depends on the ionization threshold of that state and the rate at which the voltage increases \cite{GallagherRA,Fielding}.  The rise time of the impulse determines the degree to which evolving wavepackets sample crossings nearer an adiabatic or diabatic limit, and this affects the appearance potential.   

These dynamics have been reasonably well studied for nitric oxide \cite{Fielding}, enabling us to assign features observed in the SFI spectrum to the diabatic field ionization of selected $n_0 f (2)$ states via the NO$^+$ rotational thresholds, $N^+=0$ and 2.  These appearance potentials constitute a spectrum that we can interpret to gauge the binding energies in the sample at zero field.  In previous work, we have used this spectroscopic approach to distinguish the $n_0f(2)$ NO Rydberg gas, as initially prepared, from its state a few hundred microseconds later after $\ell$-mixing and ultimately as it avalanches to a state of very low binding energy \cite{Haenel2017}.   

The initial step of $\ell$-mixing is signified by a shift of the primary $n_0f(2)$ signal to slightly higher voltage.  We can understand this to reflect collisions of surviving Rydberg molecules with a growing distribution of plasma electrons.  Migration in $\ell$ forms a Rydberg gas distribution of non-penetrating states that have the original principal quantum number and nearly the original energy \cite{Haenel2017}.  States of high orbital angular momentum populated by collisional $\ell$-mixing predissociate slowly.  Thus to noticeably increase the plasma yield by lengthening the Rydberg lifetime, mm-wave transitions to non-penetrating states must occur before electron collisions cause this redistribution of orbital angular momentum.

\subsection{Effects of mm-wave radiation on the electron density of ultracold plasmas formed by UV-UV double-resonant state-selected Rydberg gases}

\subsubsection{Relaxation dynamics of $nf(2)$ Rydberg gases}

In the absence of mm-wave radiation, we readily detect $n_0f(2)$ resonances in $\omega_2$ excitation spectra as an electron signal collected by the rising voltage of a selective field ionization ramp applied 300 ns after photoexciation.  We observe similar resonances 40 $\mu$s later in $\omega_2$ spectra of the electron signal collected after complete avalanche to plasma and propagation in the molecular beam to the detection grid.  

This suggests that dynamical processes that occur early in the avalanche of Rydberg gas to plasma compete with predissociation and act to sequester energy in the separation of positive and negative charges, even though facile pathways exist for the production of neutral atoms with high kinetic energy.  The early presence of long-lived Rydberg molecules appear to play an essential role in the formation of a stable ultracold plasma state.  

We can directly observe the electron-collisional transport of $n_0f(2)$ Rydberg gases to states of high-$\ell$ over time as a shift to higher field in the appearance potential of $n_0f(2) \rightarrow N^+ = 0$ and 2 features in the SFI spectrum \cite{Haenel2017}.  The addition of a mm-wave field tuned to resonance with $n_0f(2)  \rightarrow (n_0 + 1)g(2)$ or $(n_0 - 1)g(2)$ transition increases both the amplitude of the early time SFI spectrum and the signal associated with a long-lived ultracold plasma. 

These effects of mm-wave radiation appear for selected $n_0f(2)$ features in $\omega_2$ excitation spectra (Figures \ref{fig:n42}, \ref{fig:n39PFI} and \ref{fig:n44p}), delayed SFI spectra (Figure \ref{fig:SFI}), and in the mm-wave excitation spectra of individual $n_0f(2)$ states (Figure \ref{fig:n42mm}).  In all cases, the presence of a mm-wave field of appropriate frequency increases the electron signal extracted on resonance from the evolving plasma by about a factor of two.  

We can understand this enhancement as follows.  During the critical first few hundred nanoseconds of plasma evolution, collisional $\ell$-mixing competes with the unimolecular predissociation of nitric oxide Rydberg molecules.  As observed by Gallagher and coworkers \cite{Murgu2001}, an appropriately tuned mm-wave field drives transitions that shelve $\ell = 3$ molecules in substantially longer-lived states with $\ell = 4$.  The enhancements shown in Figures \ref{fig:n42}, \ref{fig:n39PFI}, \ref{fig:n42mm} and \ref{fig:SFI} all occur at mm-wave frequencies that conform with this assignment to $n_0f(2) \rightarrow (n_0 \pm 1)g(2)$, which conserves the NO$^+$ core rotational quantum number.  

As noted above, this stabilizing effect of the mm-wave field must occur promptly at the time of photoexcitation to gain an advantage in the competition between $\ell$-mixing and predissociation.  We can thus predict that this $\ell = 3$ to $\ell = 4$ resonant stabilization mechanism would offer less advantage if applied after this initial $\ell$-mixing period, which delayed SFI spectra show to be as short as a few hundred nanoseconds.

\subsubsection{Relaxation dynamics in Rydberg gases formed by virtue of complex resonances of lower $\ell$}

The presence of a mm-wave field with a frequency of 80.4 GHz adds new structure to the $\omega_2$ spectrum that interlopes with the $f$-series elements, $38f(2)$, $39f(2)$ and $40f(2)$.  Assignment of these new resonant features to particular principal quantum numbers depends on our assessment of their orbital angular momentum, $\ell$, and $N^+$, the ion rotational state to which they converge.  At $n=38$, the energy distance between Rydberg states of successive principal quantum number falls to a value which is smaller than the $N^+=0$ to $N^+=1$ separation between ion rotational states.  This equivalence of rotational and electronic energy causes series to overlap, creating complex resonances and an evident suppression of transition density often described as a stroboscopic region \cite{labastie1984}. 

The new features that appear in the lower frames of Figure \ref{fig:n44p} conform with the zeroth-order positions of $43p(0)$ and $41d(1)$, and $44p(0)$ and $42d(1)$, respectively.  Assuming constant quantum defects,  $\delta_p=0.7$ for $n_0p(0)$ and $\delta_d=-0.05$ for $n_0d(1)$ \cite{Jungen:2003}, we can estimate in each case that in zeroth-order these $n_0p$ and $n_0d$ states lie no more than 4 GHz apart.  Thus, here  in the region of a stroboscopic relation between the $n$ to $n+1$ Rydberg interval and the NO$^+$ $N^+ =0$ to 1 rotational spacing \cite{labastie,kay} interloping lower-$\ell$ Rydberg states can form $43p(0)/41d(1)$ and $44p(0)/42d(1)$ complex resonances.  

To within less than 1 cm$^{-1}$, the observed positions of the features added to the $\omega_2$ spectrum in Figure \ref{fig:n44p} also coincide with $41f(1)$ and $41g(1)$, and $42f(1)$ and $42g(1)$, respectively.  Thus, one might imagine that the observed bright states borrow lifetime by mixing with these underlying non-penetrating configurations \cite{Bryant:1992,Bryant:1994}, and that this field-free mixing affords a stability that permits these complex resonances to appear as a weak feature in the spectrum detected in the prompt-plasma signal collected by a rising SFI ramp.  

But, as we see from the upper frames of Figure \ref{fig:n44p}, no such resonance appears in the $\omega_2$ excitation spectrum confined to sample the Rydberg structure isolated by selective field ionization, nor does the Rydberg gas prepared in this state evolve to form the signal of a long-lived ultracold plasma.  Thus, even though the Rydberg states formed by these complex resonances have sufficient lifetime to avalanche to plasma in the core of the Rydberg gas ellipsoid, the Rydberg density in the wings must decay to neutral atoms before field ionization by an SFI ramp applied 300 ns later, and in the absence of mm-wave radiation, we see no late peak.  

The 80.4 GHz field in each case resonates with a corresponding transition from $n_0d(1)$ to a state of higher-$\ell$ built on NO$^+$, $N^+=1$, effectively shelving Rydberg molecules formed by $\omega_2$ excitation at the positions of these two complex resonances.  The weak late peaks that appear under these conditions suggest that the formation of an ultracold plasma arrested state requires the presence of an enduring population of Rydberg molecules.   

\subsubsection{Lineshapes in the mm-wave excitation spectra of state-selected $nf(2)$ Rydberg gases}

The mm-wave Rydberg-Rydberg excitation spectra shown in Figure \ref{fig:n42mm} evidence the same mm-wave-produced factor-of-two enhancement of particular resonances in the $\omega_2$ spectrum.  Pump-probe experiments measure lifetimes of $nf(2)$ states in this range no shorter than tens of nanoseconds \cite{Vrakking1995}.  Higher-$\ell$ states, including those in the $ng(2)$ series live substantially longer.  Yet the stabilizing Rydberg-Rydberg transitions defined by the lineshapes in Figure \ref{fig:n42mm} do not occur with the sharp linewidths that would correspond to lifetimes as long as these.  Instead, we see signs of transition moments broadened over intervals of about 1 GHz.  

We have determined that these resonances appear in a regime of linear absorption.  Thus, we cannot attribute the lineshapes we see to power broadening.  The presence of DC fields alter the intensities of features in the $\omega_2$ and mm-wave spectra.  We use these effects as an {\it in situ} probe to minimize the DC field to an interval of $\pm 50$ mV.  But, we observe no variation in linewidth over this range.  

To a varying extent, these spectra display slight but evident asymmetry.  We can account for this asymmetry by fitting these features to Fano profiles, described by a lineshape function \cite{Fano1961,Fano1965}:
\begin{subequations}
\label{eq:Fano}
\begin{equation}
\sigma(\epsilon)=\dfrac{(q+\epsilon)^{2}}{1+\epsilon^{2}}\label{eq:Fanoprofile} 
\end{equation}
in which the absorption cross section, $\sigma(\epsilon)$ is described in terms of an asymmetry parameter, $q$, and detuning, $\epsilon$, defined by:
\begin{eqnarray}
\epsilon =\dfrac{(\omega_{3}-E_{n})}{\Gamma/2}.\label{eq:reducedenergy}
\end{eqnarray}
\end{subequations}
where $E_n$ is the frequency of the resonance, $\Gamma$ is a linewidth, and $\omega_3$ is the scanned frequency of the mm-wave field.  Table \ref{tab:fano} lists the Fano parameters defining cross sections that, when scaled and offset, produce the lineshapes drawn through the data in Figure \ref{fig:n42mm}.  

The resonant frequencies obtained from these fits conform reasonably with values calculated for the corresponding $n_0f(2) \rightarrow (n_0 \pm 1)g(2)$ transitions adopting $\delta_f = 0.01$ and $\delta_g = 0.003$ as representative quantum defects.  

\begin{table}
 \caption{ Fano lineshape parameters describing UV-UV-mm-wave triple-resonant spectra of absorption, $n_0f(2) \rightarrow (n_0 + 1)g(2)$, and stimulated emission, $n_0f(2) \rightarrow (n_0 - 1)g(2)$, transitions evident in Figure \ref{fig:n42mm}.} 
 \label{tab:fano}
 \begin{tabular}{cccccccc}  
\toprule
 & & $\hspace{-10 pt} (n_0 + 1)g(2)$ & &  &   & $\hspace{-10 pt} (n_0 - 1)g(2)$  \\
 $n_0$  & \hspace{0 pt} $\omega_0$ (GHz) & $\Gamma$ (GHz) & $q$ & &$\omega_0$ (GHz) & $\Gamma$ (GHz) & $q$ \\
  \hline 
  $39f(2)$ &  108.6 &  0.7 & -20 & \hspace{0 pt}  \hspace{0 pt} & 113.5 & 0.7 & 20 \\
  $42f(2)$ &  87.1 &  1.2 & -4 & \hspace{0 pt}  \hspace{0 pt} & 90.5 & 1.2 & -4 \\
  $43f(2)$ &  81.3 &  1.0 & -8 & \hspace{0 pt}  \hspace{0 pt} & 84.3 & 1.0 & -8 \\
  $45f(2)$ &  71.0 &  1.1 & -5 & \hspace{0 pt}  \hspace{0 pt} & 73.5 & 1.1 & 5 \\
 \hline
\end{tabular}
\end{table}

Fano lineshapes arise when an electromagnetic field drives transitions from a common originating state to a discrete excited state and directly to a continuum to which the discrete state is coupled.  These transition moments interfere, giving rise to a lineshape described by Eq \ref{eq:Fanoprofile} that spans a coupling width $\Gamma$.

We know from measured predissociation lifetimes that the coupling of $n_0f(2)$ and $(n_0 \pm 1)g(2)$ states with dissociation continua give rise to linewidths much narrower than 1 GHz.  But, in addition to this weak coupling with neutral continua, each discrete $n_0f(2)$ and $(n_0 \pm 1)g(2)$ level is embedded in a quasi-continuum background of states formed by other Rydberg molecules in the evolving plasma with varying values of zeroth-order quantum numbers, $n$, $\ell$ and $N^+$, coupled by dipole-dipole interactions that conserve total angular momentum and energy to within a coupling width, $\Gamma$.  

The width of a resonant interaction between a pair of NO molecules $i$ and $j$ varies with distance as $t_{ij}/r_{ij}^3$, where $t_{ij}$ is the dipole-dipole matrix element for the interaction.  Dipole-dipole interactions serve to distribute the zeroth-order $n_0f(2) \rightarrow (n_0 \pm 1)g(2)$ character of the bright-state transition over initial and final states of the aggregated system -- so as to include the contribution of neighbouring molecules with other combinations of $n$, $\ell$ and $N^+$ that fall within the coupling width, $\Gamma$.  Evidently, the conditions of our experiment and the principal quantum numbers we have chosen to investigate dictate a coupling width in the neighbourhood of 1 GHz.

\section{Conclusions}

Resonant mm-wave fields drive $n_0f(2) \rightarrow (n_0 \pm 1)g(2)$ transitions in a state-selected $n_0f(2)$ Rydberg gas of NO.  This transformation, which produces a clear signature in the SFI spectrum, dramatically increases the early time intensity of high-Rydberg resonances in the SFI-detected $\omega_2$ spectrum.  We associate these enhanced features with a decrease in the rate of predissociation owing to an increase in Rydberg orbital angular momentum.  This same signature of early time Rydberg stabilization appears as an enhanced long-time plasma signal, extracted as a late peak after 40 $\mu$s of field-free evolution.  

In addition to its single dominant $n_0f(2)$ series, the $\omega_2$ spectrum of NO exhibits interloping resonances that appear at the positions of $44p(0) - 43d(1)$ and  $43p(0) - 42d(1)$ complex resonances.  In the absence of a mm-wave field, these states give rise to features in an $\omega_2$ excitation spectrum when we collect the initial signal extracted at low voltage in the beginning of an SFI ramp applied a few hundred nanoseconds after excitation.  However, this SFI spectrum shows no field-ionization signal of surviving Rydberg molecules in these states, and no late-peak at this position in the $\omega_2$ excitation spectrum.  

With the addition of a suitable mm-wave field, these complex resonances produce a strong SFI signal of Rydberg molecules that survives for hundreds of nanoseconds and interloping late peaks.  We conclude from this that avalanche alone does not guarantee a plasma state of arrested relaxation.  But rather, the formation of an arrested phase requires both avalanche-produced NO$^+$ ions and a persistent population of long-lived Rydberg molecules.  

\section*{Acknowledgments}
This work was supported by the US Air Force Office of Scientific Research (Grant No. FA9550-17-1-0343), together with the Natural Sciences and Engineering Research Council of Canada (NSERC), the Canada Foundation for Innovation (CFI) and the British Columbia Knowledge Development Fund (BCKDF). 

\bibliography{MWpaper.bib}

\end{document}